\documentclass[journal=nalefd,manuscript=letter]{achemso}
\pdfoutput=1
\usepackage[version=3]{mhchem} % Formula subscripts using \ce{}
\usepackage[T1]{fontenc}       % Use modern font encodings

\author{Jakob Kammhuber}
\author{Maja C. Cassidy}
\author{Hao Zhang}
\author{{\"O}nder G{\"u}l}
\author{Fei Pei}
\author{Michiel W. A. de Moor}
\author{Bas Nijholt}
\affiliation[]
{QuTech and Kavli Institute of Nanoscience, Delft University of Technology, 2628 CJ Delft, The Netherlands}

\author{Kenji Watanabe}
\author{Takashi Taniguchi}
\affiliation[]
{Advanced Materials Laboratory, National Institute for Materials Science, 1-1 Namiki, Tsukuba, 305-0044, Japan}

\author{Diana Car}
\affiliation[]
{Department of Applied Physics, Eindhoven University of Technology, 5600 MB Eindhoven, The Netherlands}

\author{S\'ebastien R. Plissard}
\affiliation
{QuTech and Kavli Institute of Nanoscience, Delft University of Technology, 2628 CJ Delft, The Netherlands}
\alsoaffiliation
{Department of Applied Physics, Eindhoven University of Technology, 5600 MB Eindhoven, The Netherlands}
\alsoaffiliation
[Current address: CNRS-Laboratoire d'Analyse et d'Architecture des Systemes (LAAS), Universit\'e de Toulouse, 7 avenue du Colonel Roche, F-31400 Toulouse, France]
{CNRS-Laboratoire d'Analyse et d'Architecture des Systemes (LAAS), Universit\'e de Toulouse, 7 avenue du Colonel Roche, F-31400 Toulouse, France}

\author{Erik P. A. M. Bakkers}
\affiliation
{QuTech and Kavli Institute of Nanoscience, Delft University of Technology, 2628 CJ Delft, The Netherlands}
\alsoaffiliation
{Department of Applied Physics, Eindhoven University of Technology, 5600 MB Eindhoven, The Netherlands}

\author{Leo P. Kouwenhoven}
\affiliation
{QuTech and Kavli Institute of Nanoscience, Delft University of Technology, 2628 CJ Delft, The Netherlands}
\email{l.p.kouwenhoven@tudelft.nl}

\title[QPC Nanoletters]
  {Conductance Quantization at zero magnetic field in InSb nanowires}

\keywords{Quantum point contact, conductance quantization, nanowire, InSb, subband, orbital effects\\}

\begin{document}
\begin{tocentry}
 \includegraphics{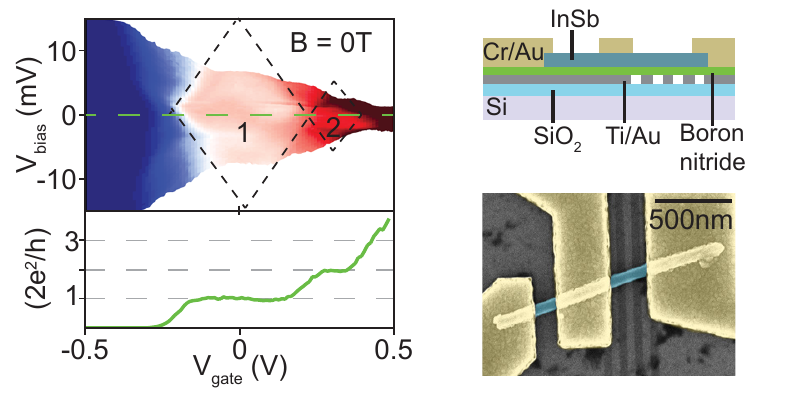}
\end{tocentry}

\pagebreak
\begin{abstract}

Ballistic electron transport is a key requirement for existence of a topological phase transition in proximitized InSb nanowires.
However, measurements of quantized conductance as direct evidence of ballistic transport have so far been obscured due to
the increased chance of backscattering in one dimensional nanowires. We show that by 
improving the nanowire-metal interface as well as the dielectric environment we can consistently achieve conductance 
quantization at zero magnetic field. Additionally, studying the sub-band evolution in a rotating magnetic field reveals an orbital degeneracy between
the second and third sub-bands for perpendicular fields above $1\,\text{T}$.

\end{abstract}


Semiconducting nanowires made from InAs and InSb are prime candidates for the investigation of novel phenomena in electronic devices. The intrinsic strong spin-orbit interaction (SOI) and large g-factor combined with flexible fabrication has resulted in these materials being investigated for applications in quantum computing,\cite{nadj2010spin}\textsuperscript{,}\cite{vandenBerg2013fast}\textsuperscript{,}\cite{petersson2012circuit} spintronics\cite{Liang2012nanolett}\textsuperscript{,}\cite{rossella2014nanoscale}\textsuperscript{,}\cite{Zutic2004RevModPhys} and Cooper pair splitters.\cite{hofstetter2009cooper}\textsuperscript{,}\cite{das2012high} More recently, these nanowires have been investigated as solid-state hosts for Majorana zero modes (MZMs).
\cite{mourik2012signatures}\textsuperscript{,}\cite{deng2012anomalous}\textsuperscript{,}\cite{churchill2013PRB}\textsuperscript{,}\cite{deng2014parity} By bringing a one dimensional (1D) nanowire with strong SOI into close contact with a superconductor under an external magnetic field, a region with inverted band structure emerges, creating MZMs at its ends. 
Together with strong SOI and induced superconductivity, a key requirement for MZMs is quasi-ballistic electron transport along the length of the proximitized region in the nanowire, with a controlled odd number of occupied modes.\cite{Lutchyn2011PRL} In the absence of scattering, the motion of 1D confined electrons will be restricted to discrete energy bands resulting in quantized conductance plateaus.\cite{vanWees1988PRL}\textsuperscript{,}\cite{Wharam1988JPC}
Measurements of quantized conductance in the nanowires therefore provide direct evidence for controlled mode occupation, as well as ballistic transport in these nanowires.

Although now routine in gate defined quantum point contacts (QPC) in two-dimensional electron gases (2DEG), \cite{vanWees1988PRL}\textsuperscript{,}\cite{Wharam1988JPC}\textsuperscript{,}\cite{Chou2008APL}\textsuperscript{,}\cite{Tobben19995SST}\textsuperscript{,}\cite{Koester1996PRB}
quantized conductance in one dimensional semiconductor nanowires is more difficult to achieve. In a 1D nanowire, scattering events along the electrons path to and through the constriction between the source 
and drain contacts have an increased probability of reflection, obscuring the observation of quantized conductance.\cite{vanweperen2012quantized} These scattering events may be due to impurities 
and imperfections in the crystal lattice, or due to surface states that create inhomogeneities in the local electrostatic environment.\cite{Gul2014nanotechnology} A Schottky barrier between the 
nanowire and metallic contacts will result in additional backscattering events, further smearing out the quantized conductance plateaus. To date, quantized conductance in InSb nanowires has only 
been observed at high magnetic fields ($>\,4\,\text{T}$), where electron backscattering is strongly suppressed.\cite{vanweperen2012quantized} No quantization has been observed in InSb for 
magnetic fields below $1\,\text{T}$, where the topological transition is expected to take place.\cite{mourik2012signatures}      

Here we demonstrate conductance quantization in InSb nanowires at zero magnetic field. We have developed a robust fabrication recipe for observing quantized conductance by optimizing both the metal-nanowire contact interface and dielectric environment through the use of hexagonal boron nitride (hBN) as a gate dielectric. We study the evolution of the quantized conductance plateaus with both source-drain bias as well as magnetic field, and extract values for the Land\'e g-factor of the first three sub-bands in the nanowire. Additionally, we observe an orbital energy degeneracy of the second and third sub-bands at finite magnetic fields applied perpendicular to the nanowire.

\begin{figure}
\includegraphics{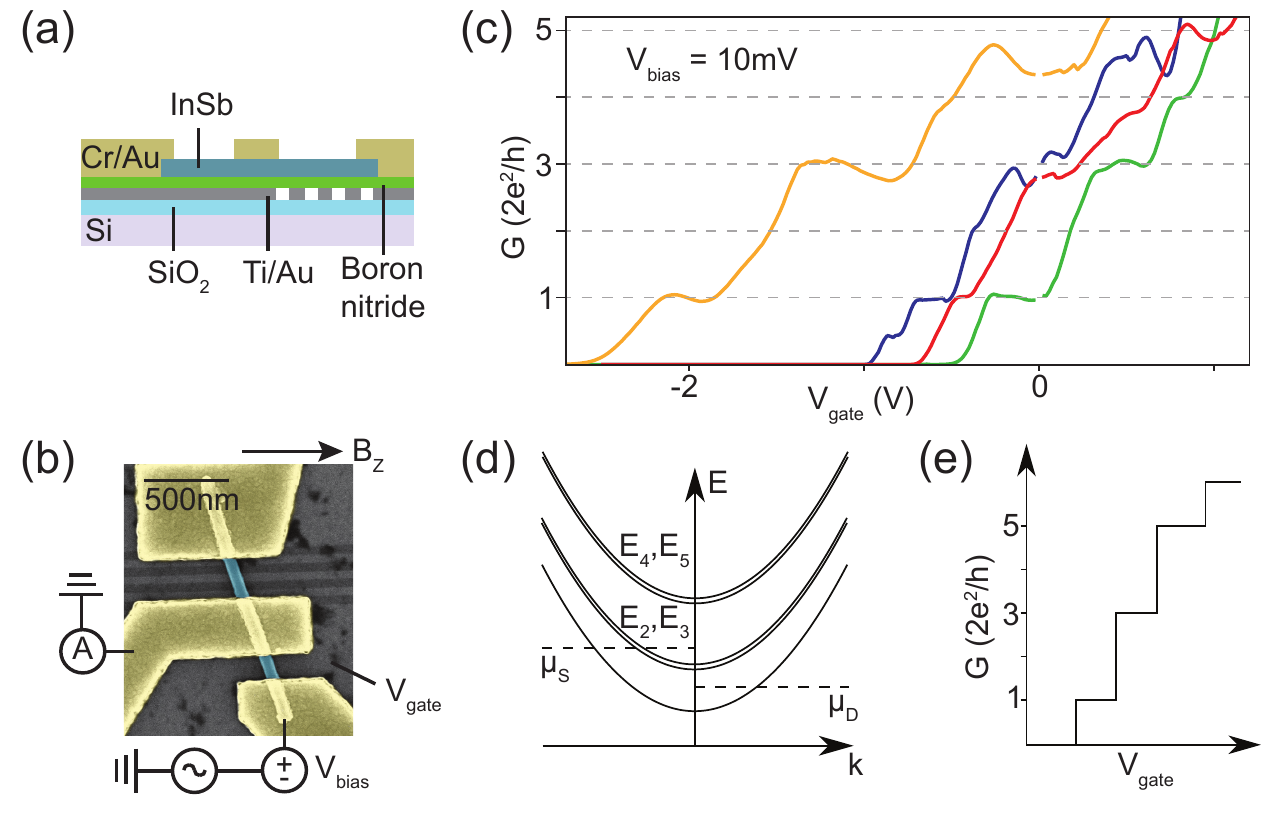}
  \caption{(a) Cross-sectional schematic and (b) false color SEM image of a typical device. An InSb nanowire (blue) contacted by Cr/Au (yellow) is 
  deposited on Ti/Au metal gates (grey) covered with hexagonal boron nitride (green) as insulating dielectric.
  (c) Pinch-off traces of four different devices each showing quantized conductance plateaus at high bias voltage ($\text{V}_{\text{bias}} = 10\,\text{mV}$).
  (d) Schematic diagram of the first five sub-bands in a nanowire. At zero magnetic field, each spin-degenerate sub-band below the Fermi level
  contributes a conductance of $\text{G}_0=2\text{e}^2/\text{h}$. Due to the rotational symmetry of the nanowires $\text{E}_2$,$\text{E}_3$
  and $\text{E}_4$,$\text{E}_5$ are almost degenerate.
  (e) Sketch of the expected conductance steps as function of $\text{V}_{\text{gate}}$ at high bias voltage showing suppression of the second and fourth 
  plateaus due to orbital sub-band degeneracy.}
  \label{fgr:2}
\end{figure}

Figure \ref{fgr:2}a) shows a cross-sectional view of our devices. They consist of an intrinsic Si-substrate with local metallic gates made of Ti/Au ($5$/$10\text{nm}$), 
on top of which a sheet of hexagonal boron nitride (hBN) is mechanically transferred as the dielectric. 
The chemical stability, atomic flatness, and high breakdown voltage\cite{dean2010boron}, together with the well established dry transfer mechanism \cite{castellanos20142Dmaterials}
makes hBN an ideal dielectric for our nanowire devices. InSb nanowires grown by metal-organic vapor phase epitaxy \cite{plissard2012insb}\textsuperscript{,}\cite{car2014rationally} 
($1$ - $3 \mu\text{m}$ long and $ 70$ - $90\text{nm}$ diameter) are transferred deterministically with a micro-manipulator\cite{flohr2011manipulating} onto the hBN dielectric.
Electrical contacts to the nanowire (evaporated Cr/Au ($10/100\,\text{nm}$), $150$ - $400$nm spacing) are defined by electron beam lithography.
Before contact deposition, the surface oxide of the nanowires is removed using sulfur passivation\cite{suyatin2007Nanotechnology} followed by a short in situ He-ion mill.
Residual sulfur from the passivation step also induces surface doping, which aids contact 
transparency. Further details of the fabrication are included in the supporting information.
A top view scanning electron microscope image of a finished device is shown in Fig \ref{fgr:2}b). The samples are mounted in a 
dilution refrigerator with a base temperature of $15\,\text{mK}$ and measured using standard lock-in techniques at $73$ Hz with an excitation V$_{RMS} = 70\,\mu\text{V}$. 
Voltage is applied to the outer contact and current measured through the grounded central contact, while the third, unused contact is left floating.

We first characterize each device by sweeping the voltage on the underlying gate $\text{V}_{ \text{gate} }$ at fixed bias voltage $\text{V}_{ \text{bias} } = 10\,\text{mV}$
across the sample. Conductance is obtained directly from the measured current $\text{G} = \text{I}/\text{V}_{ \text{bias} }$ and an appropriate series resistance is 
subtracted in each case (see supporting information). 
Figure \ref{fgr:2}c) plots the conductance of the nanowire as function of gate voltage for four different devices fabricated on the same chip. 
Devices with both fine gates as well as wide back gates have been measured. We find that while fine gates allow more flexible gating, devices with wide back gates showed more 
pronounced conductance plateaus even after extensive tuning of the fine gates. Data from additional devices all fabricated on the same chip is included in the 
supporting information.

As seen in Fig \ref{fgr:2}c) all devices show well defined plateaus at $\text{G}_0$ and $3\,\text{G}_0$ but the plateaus at $2\,\text{G}_0$ and $4\,\text{G}_0$
appear smaller or even completely absent. Unlike QPC's formed in 2DEGs, nanowires possess rotational symmetry. This symmetry can give rise to additional orbital degeneracies 
in the energies for the 2\textsuperscript{nd} and 3\textsuperscript{rd} as well as the 4\textsuperscript{th} and 5\textsuperscript{th} sub-band 
(Fig \ref{fgr:2}d).\cite{ford2012observation}\textsuperscript{,}\cite{vanweperen2014thesis}
In conductance measurements at finite bias, sub-bands that are close in energy or degenerate will be populated at similar values in gate voltage giving a double step of 
$4\, \frac{e^2}{h}$ instead of $2\, \frac{e^2}{h}$, which explains the suppressed plateaus at $2$ and $4\, \text{G}_0$ (Fig \ref{fgr:2}e).\cite{krans1995QPCmetallic}

\begin{figure}
\includegraphics{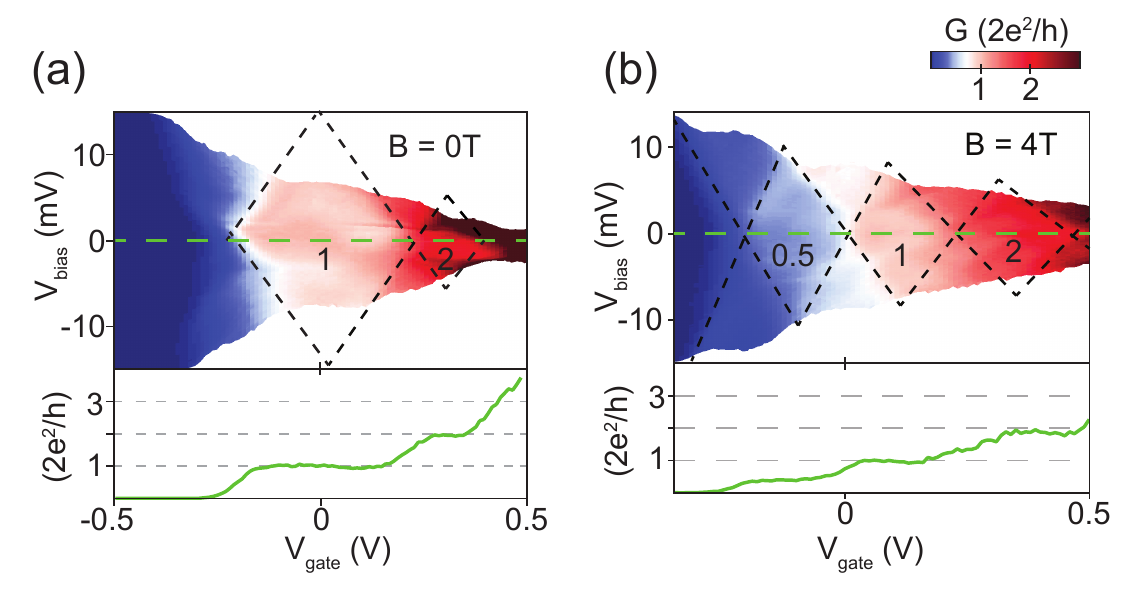}
  \caption{Color-plot of the differential conductance $\text{G} = \text{dI}/\text{dV}_\text{bias}$ as function of $\text{V}_{\text{bias}}$ and $\text{V}_{\text{gate}}$ at (a) $\text{B} = 0\,\text{T}$ and (b) $\text{B}_Z = 4\,\text{T}$. A line cut along zero bias voltage is shown in the bottom panel.
  Plateaus appear as diamonds and are indicated by black dotted lines.
  }
  \label{fgr:3}
\end{figure}

To investigate this phenomenon in more detail, we measure the differential conductance $\text{G} = \text{dI}/\text{dV}_\text{bias}$ as function of gate voltage and bias voltage
for one of these devices (corresponding to the green trace in Fig \ref{fgr:2}c). This data is shown in Fig \ref{fgr:3}a) as a color plot, with a line cut along zero bias voltage added in the bottom panel. 
At zero bias voltage an extended plateau is visible at $1\, \text{G}_0$, together
with an additional small plateau at $2\, \text{G}_0$ which was not visible in the linear conductance data of Figure \ref{fgr:2}c). The existence of this small $2\,\text{G}_0$ plateau
indicates that the device has a small, but finite energy splitting between the second and third sub-band which was not resolved at high bias.
At finite bias voltage the conductance will only be quantized in integer values of $\text{G}_0$ if both $\mu_{\text{source}}$ and $\mu_{\text{drain}}$ occupy the same sub-band. 
This creates diamond shaped regions of constant conductance indicated by black dotted lines in Fig \ref{fgr:3}a). At the tip of the diamond the two dotted lines cross when 
$\text{V}_{\text{bias}}$ is equal to the sub-band energy spacing $\Delta \text{E}_{\text{subband}}$. From this we extract $\Delta \text{E}_{\text{subband}}$ and the lever-arm 
$\eta$ of the bottom gate via $\eta\,\text{V}_{\text{gate}} = \Delta \text{E}_{subband}$.\cite{Kouwenhoven1989PRB}\textsuperscript{,}
A finite magnetic field breaks time reversal symmetry, lifting spin degeneracy and splitting the individual spin sub-bands $\text{E}_{n,\uparrow / \downarrow}$ by the 
Zeeman energy $\text{E}_{\text{Zeeman}} = g \mu_B \text{B}$.
Here $\mu_B$ denotes the Bohr magneton and g the Land\'e g-factor.  Experimentally this splitting manifests as the appearance of additional half integer steps $\frac{N}{2} \cdot 2e^2/h$.
At $\text{B}=4\,\text{T}$ we clearly observe this for the first sub-band as shown in Fig \ref{fgr:3}b) where an additional plateau emerges at $0.5\,\text{G}_0$.
Similarly, the second sub-band should also split into two plateaus at $1.5$ and $2\,\text{G}_0$. However only the $2\,\text{G}_0$ plateau is visible, 
suggesting that the orbital degeneracy between $\text{E}_{2,\uparrow}$ and $\text{E}_{3,\uparrow}$ remains at finite magnetic field.

\begin{figure}
\includegraphics{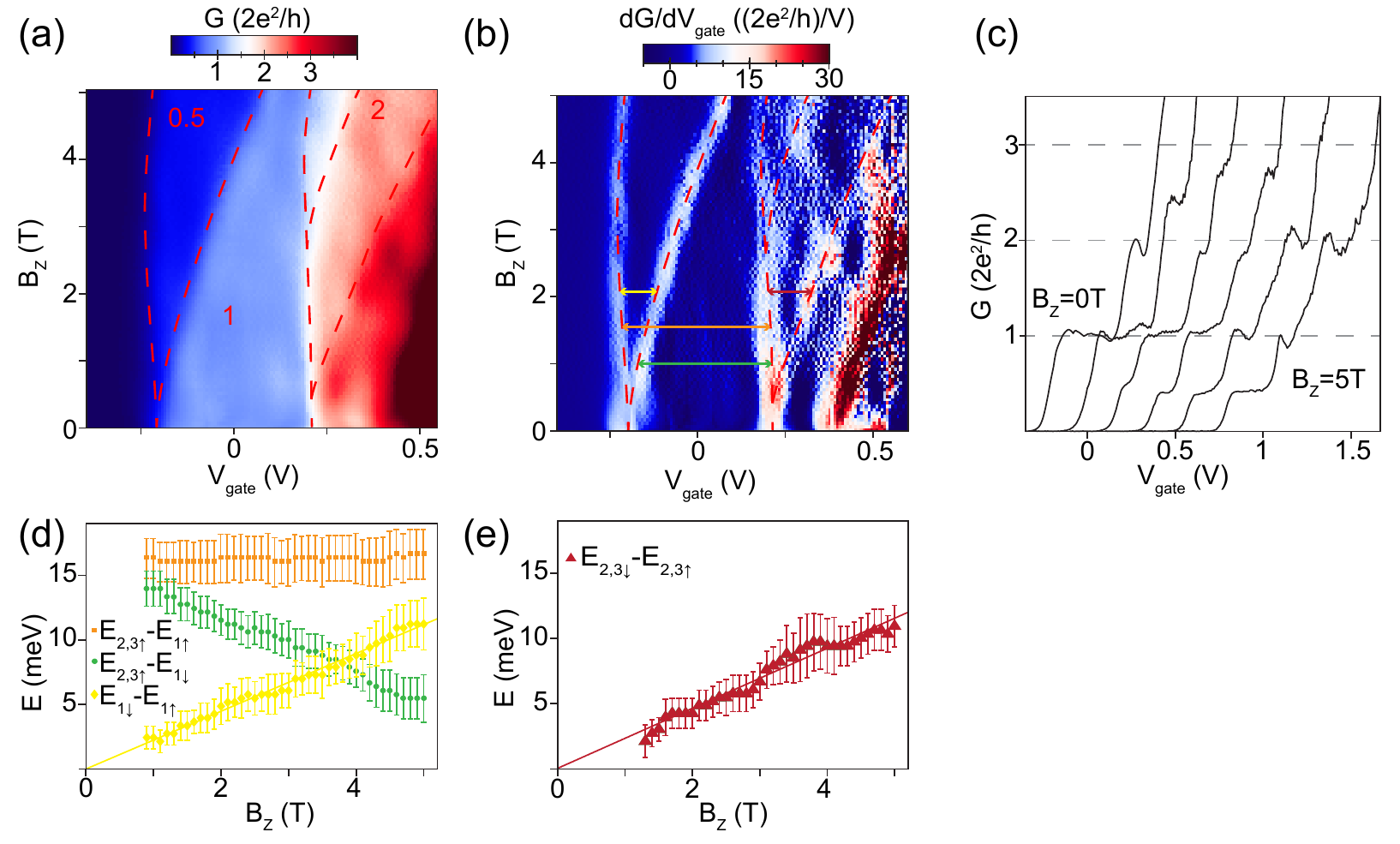}
  \caption{(a) Differential conductance $\text{G} = \text{dI}/\text{dV}_{\text{bias}}$ and (b) transconductance $ \text{dG}/ \text{dV}_\text{gate}$
  as function of magnetic field along $\text{B}_\text{Z}$ and $\text{V}_{\text{gate}}$ taken at $\text{V}_{bias} = 0\,\text{mV}$. The level spacings 
  plotted in d,e) are marked by arrows of corresponding color.
  Red dashed lines indicating the sub-band spacing in a,b) are drawn as guide to the eye.
  (c) Linecuts of a) in steps of 1T and offset by $200 \text{mV}$ for clarity.
  (d) Energy level spacings of $\text{E}_{1\downarrow}-\text{E}_{1\uparrow}$ (yellow), $\text{E}_{2,3\uparrow}-\text{E}_{1\downarrow}$ (green) and 
  $\text{E}_{2,3\uparrow}-\text{E}_{1\uparrow}$ (orange) extracted from the $0.5$ and $1\text{G}_0$ plateau in (a). A linear fit to 
  $\text{E}_{1\downarrow}-\text{E}_{1\uparrow}$ fixed at the origin gives the g-factor of the first sub-band $\text{g}_1=39 \pm 1$. 
  (e) Energy spacing of $\text{E}_{2,3\downarrow}-\text{E}_{2,3\uparrow}$ extracted from the $2\text{G}_0$ plateau with $\text{g}_{2,3} = 40\pm 1$.
  }
  \label{fgr:4}
\end{figure}

The full evolution in magnetic field of the conductance and transconductance is shown in Fig \ref{fgr:4}a,b) and individual line traces of the conductance taken in steps of 
$1\,\text{T}$ are presented in Fig \ref{fgr:4}c). While the plateau at $1\,\text{G}_0$ remains very flat up to high magnetic fields, the second plateau at $2\,\text{G}_0$ increases in 
height for magnetic fields larger than $400\,\text{mT}$.
Around $1\,\text{T}$ two new plateaus emerge with similar slope at $0.5$ and $2\,\text{G}_0$. These correspond to the lower energy spin sub-bands $\text{E}_{1\uparrow}$
and $\text{E}_{2,3\uparrow}$. Here we can clearly see experimentally that the non-degenerate orbital state at zero field transforms into a degenerate orbital state at finite field and 
that $\text{E}_{2,3\uparrow}$ remain degenerate over a magnetic field range of several Tesla. 

From the individual gate traces we convert the plateau width to energy by using the lever arm $\eta$ extracted from Figure \ref{fgr:3}. This way we can directly extract the 
sub-band spacing $\text{E}_{2\uparrow}-\text{E}_{1\uparrow}$ and the individual g-factors $\text{g}_1$, $\text{g}_{2,3}$ through a linear fit fixed at the origin to 
$\text{E}_{1\downarrow}-\text{E}_{1\uparrow}$ and $\text{E}_{2,3\downarrow}-\text{E}_{2,3\uparrow}$. We find $\text{g}_1 = 39\pm 1$ $\text{g}_{2,3} = 40 \pm 1$
and a constant sub-band spacing $\text{E}_{2\uparrow}-\text{E}_{1\uparrow} \approx 16\,meV$.

\begin{figure}
\includegraphics{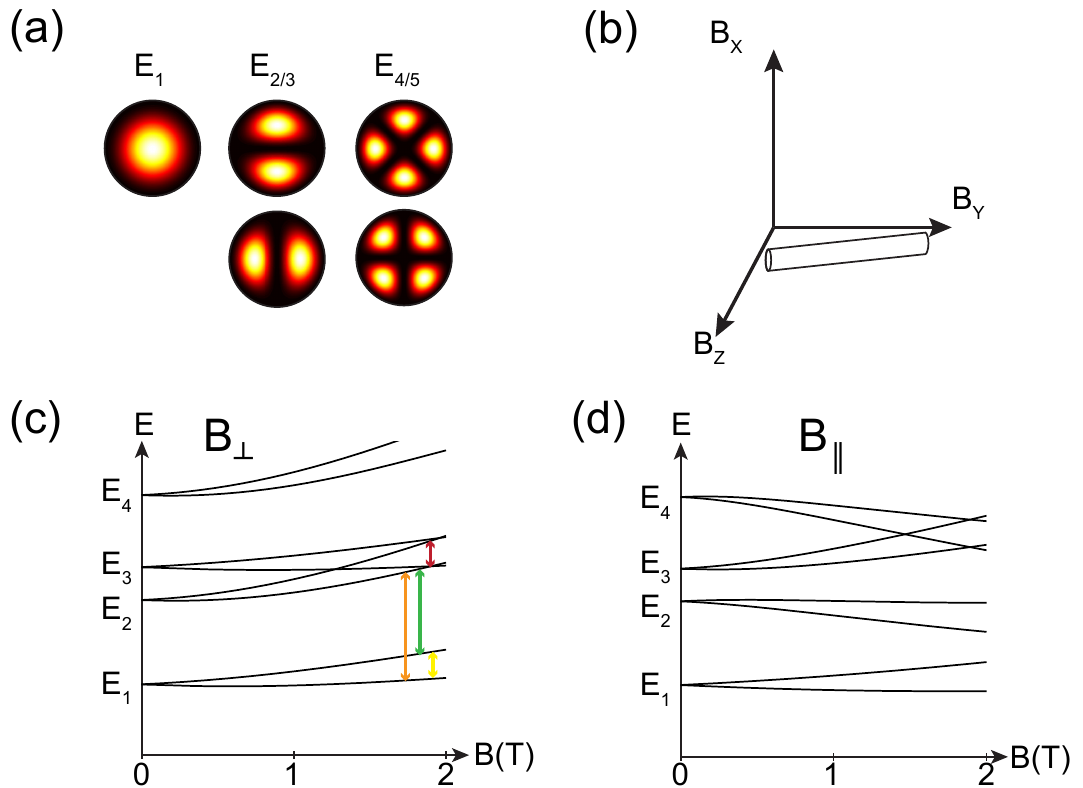}
  \caption{ 
  (a) Probability density of the first 5 sub-bands of a cylindrical nanowire.
  (b) Orientation of the nanowire with respect to the magnetic field axes. 
  (c,d) Numerical simulations of the sub-band dispersion of a InSb nanowire in perpendicular (c) and parallel (d) magnetic field.
  }
  \label{fgr:5}
\end{figure}

\begin{figure}
\includegraphics{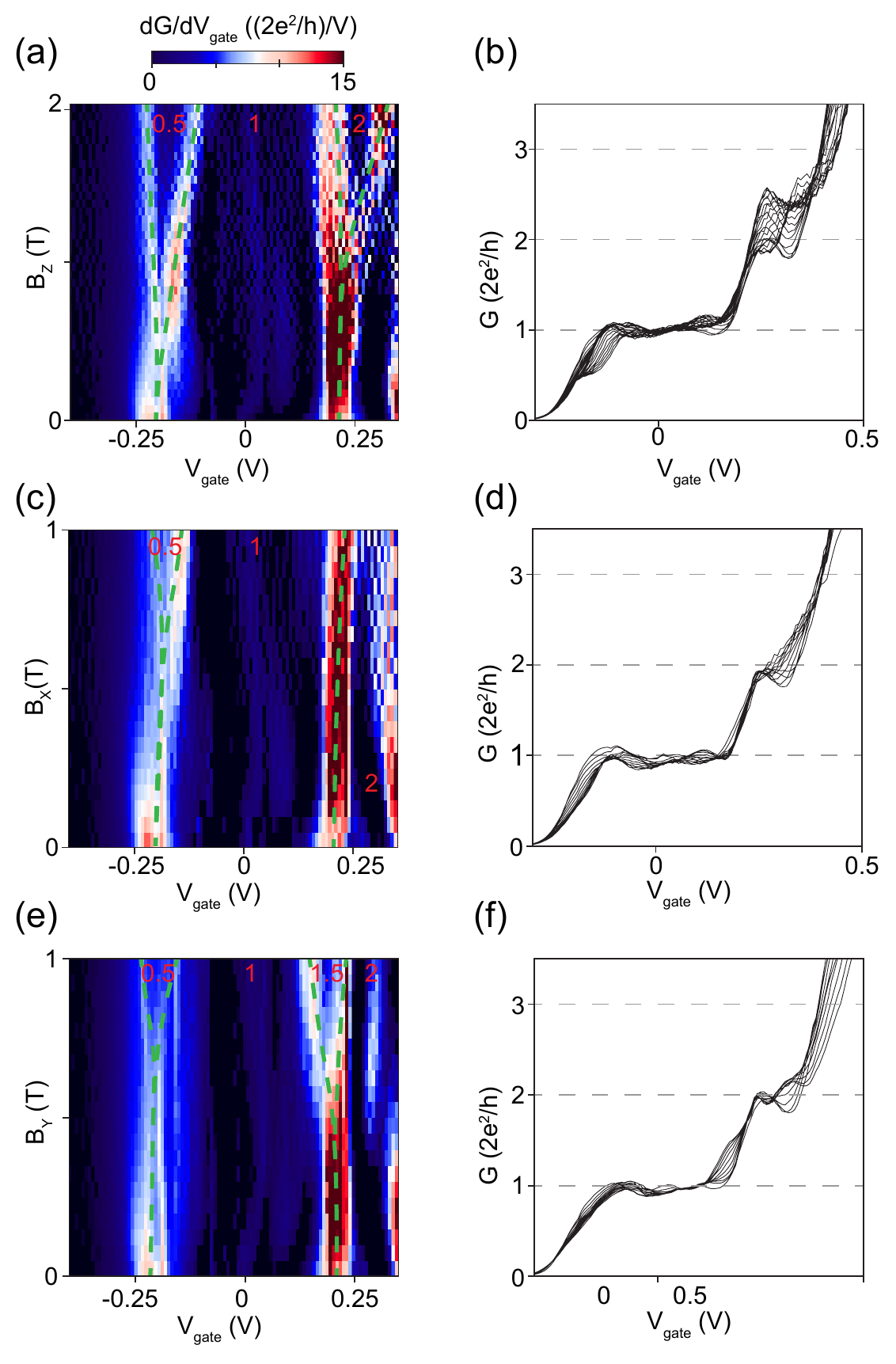}
  \caption{ Transconductance $\text{dG}/ \text{dV}_\text{gate}$ and differential conductance G for three different directions of the magnetic field all taken at $\text{V}_{bias}=0\,\text{mV}$.
  Green dashed lines indicating the sub-band spacing in a,c,e) are drawn as guide to the eye and red numbers label the height of the conductance plateaus. $\text{B}_\text{Z}$ is increased from $0-2 \,\text{T}$ and $\text{B}_\text{X,Y}$ from $0-1\,\text{T}$
   (a,b) Magnetic field aligned along $\text{B}_\text{Z}$.
  (c,d) Magnetic field aligned along $\text{B}_\text{X}$.
  (e,f) Magnetic field aligned along $\text{B}_\text{Y}$.
  }
  \label{fgr:6}
\end{figure}

Orbital degeneracy of sub-bands has previously been observed in metallic point contacts\cite{krans1995QPCmetallic} and recently also
in passivated narrow InAs nanowires with highly symmetric conducting channels\cite{ford2012observation}. However, the magnetoconductance of InSb nanowires 
may deviate significantly from the results found in InAs nanowires. In InAs, Fermi level pinning leads to conduction close to the nanowire surface
\cite{Scheffler2009APL}\textsuperscript{,}\cite{Halpern2012APL} which strongly influences the sub-band dispersion in magnetic field.\cite{Holloway2015PRB}\textsuperscript{,}\cite{Tserkovnyak2006PRB}
InSb has no surface accumulation\cite{King2008JAP} and the electron wave-function will be more strongly confined in the center of the nanowire.
For cylindrical nanowires individual sub-band wave functions are given by Bessel functions with different orbital angular momentum along the wire (Fig \ref{fgr:5}a), and
numerical simulations of wires with a hexagonal cross-section show qualitatively similar results.\cite{vanweperen2014thesis}\textsuperscript{,}\cite{Vuik2015arxiv} 
An additional magnetic field will
add Zeeman splitting, but also causes orbital effects which can substantially change the sub-band dispersion depending on the orientation of the field with respect to the 
nanowires axis.\cite{Nijholt2015arxiv}
Numerical simulations of nanowires in a magnetic field show that orbital effects strongly depend on the magnetic field orientation and can dominate the 
sub-band dispersion, leading to a decrease of the energy splitting between $\text{E}_2$ and $\text{E}_3$. \cite{Nijholt2015arxiv} Furthermore these simulations also show that the orbital effects can strongly influence the phase diagram of MZMs.
\cite{Nijholt2015arxiv}
  Using the model of ref. \citenum{Nijholt2015arxiv} with the parameters of our device (wire radius: $35\,\text{nm}$; g-factor: $40$)
we simulate this change in the sub-band dispersion for a magnetic field perpendicular (Fig \ref{fgr:5}c) and parallel (Fig \ref{fgr:5}d) to the nanowire.
A perpendicular magnetic field causes $\text{E}_{2,\uparrow}$ and $\text{E}_{3,\uparrow}$ to shift higher and closer in energy, while a parallel magnetic field increases the energy splitting of the higher sub-bands $\text{E}_2$ and $\text{E}_3$, due to their different orbital angular momentum. 

Experimentally we test this by rotating the direction of the magnetic field, as shown in Fig \ref{fgr:6}. When aligning the magnetic field along $\text{B}_\text{Z}$ (almost perpendicular to
the nanowire, Fig \ref{fgr:6}a,b), a small splitting appears at the beginning of the first plateau for fields above $\text{B}_\text{Z}=0.6\,\text{T}$,
marking the onset of the $0.5$-plateau. In contrast, in the second plateau the splitting only starts above $1\,\text{T}$ and the line-cuts (Fig \ref{fgr:6}b) 
show that the new plateau emerges at $2\,\text{G}_0$. Similarly, for a magnetic field along $\text{B}_\text{X}$ (Fig \ref{fgr:6}c,d), a new plateau emerges 
around $\text{B}_\text{X}=0.6\,\text{T}$ in the first step but not in the second.
However, for the magnetic field aligned along $\text{B}_Y$ (mostly parallel to the nanowire) shown in Fig \ref{fgr:6}e,f) we do see a clear difference. 
Now two new plateaus emerge almost simultaneously around $\text{B}_Y \approx 0.75\,\text{T}$, with the second plateau at $1.5$ and not at $2\,\text{G}_0$, in agreement with the expected behavior due to orbital effects.

In conclusion we achieved substantial improvements in electrical transport measurements of InSb nanowires by using a high quality hBN dielectric and 
clearly demonstrated conductance quantization at zero magnetic field, as well as degenerate sub-bands at magnetic field above $1\,\text{T}$.
In the future these, improvements will allow the more detailed investigation of features in the $1$\textsuperscript{st} plateau, such as
signatures of a helical gap, \cite{streda2003PRL}\textsuperscript{,}\cite{pershin2004PRB} or the presence of a 0.7 anomaly.\cite{Thomas1996PRL}\textsuperscript{,}\cite{bauer2013nature}\textsuperscript{,}\cite{iqbal2013nature}
The large SOI in our InSb nanowire strongly influences the electron dispersion relation and the tunability with magnetic field could add new insight into the 
underlying physics\cite{Goulko2014PRL}. We did not see any clear features related to the $0.7$-anomaly in our devices. However, the $0.7$ state becomes more 
pronounced at higher temperatures\cite{Thomas1996PRL}. A more detailed study of the temperature dependence of conductance quantization may reveal more information about the existence
of this intriguing state in nanowire QPCs.


\begin{acknowledgement}
The authors thank M. Wimmer, P. Kim and A. Akhmerov for helpful discussions, S. Goswami for help with the hBN transfer, and D. van Woerkom for help with nanowire deposition.
This work has been supported by funding from the Marie Curie ITN S$^3$Nano, the ERC starting grant STATOPINS 638760, NWO/FOM and Microsoft Corporation Station Q.
\end{acknowledgement}

\begin{suppinfo}
The supporting information contain a detailed fabrication recipe, a discussion of the subtracted series resistance, additional data of the main device as well
as data of QPC devices fabricated with a SiO2 dielectric.

\end{suppinfo}

\bibliography{references}

\end{document}